\documentstyle[epsfig]{aipproc}
\def\del{{\partial}}
\def\bv{{\bf v}}
\def\bB{{\bf B}}
\def\<{{\langle}}
\def\>{{\rangle}}
\def\bnabla{{\bf \nabla}}

\begin{document}

\title{Numerical Models of Accretion Disks}
 
\author{Charles F. Gammie}
\address{Harvard-Smithsonian Center for Astrophysics \\
60 Garden St., MS-51, Cambridge, MA 02138 USA \\
cgammie@cfa.harvard.edu}

\maketitle

\begin{abstract}
I review recent numerical studies of accretion disks, focusing on
measurements of the turbulent shear stress, or $\alpha$, in the
shearing box model.  I conclude with a list of astronomically relevant
open questions that can be settled via future numerical experiments.

\end{abstract}

\section*{Introduction}

The last decade has brought a flood of cheap, fast workstations and
easy access to supercomputers to the astronomical community.  For those
struggling to understand complicated phenomena like the active nucleus
of NGC 4258, with its beautifully precise maser spots orbiting in a
thin, warped disk, these machines seem to offer the seductive
possibility of creating a detailed and comprehensive simulation.  One
could imagine that by including all the relevant physics, such a
simulation would self-consistently generate X-rays from the
relativistic accretion flow in the neighborhood of the black hole
(ending the debate between proponents of ADAF and disk corona models);
it would warp straightaway (giving away the origin of the warp); it
would include a chemical network to calculate the local abundance and
excitation of water; it would perform three dimensional (3D) radiative
transfer in the water maser lines; and of course, it would produce a
jet.  Our curiosity would be completely satisfied.

Of course, such global and physically complete simulations are mere
fantasy.  They are not possible now, nor will they be except in some
distant, post-Moore's-law future world.  It is not even clear that they
would be desirable!  Even if one could determine the right initial
conditions--- probably impossible even in principle-- the output would
be so complicated (cf. \cite{o85} on spiral structure) that it would
not be clear what was important.

Absent global, {\it ab initio} simulations one is reduced to solving
simplified model equations that rely on reasonable but untested
physical assumptions.  Thus models of dwarf novae (see, e.g.
\cite{c93,hmdlh98}) typically evolve an azimuthally averaged and
height-integrated set of equations for the surface density $\Sigma$:
\begin{equation}\label{ESURF}
\del_t \Sigma(r,t) = {2\over{r}}\del_r\left({1\over{r\Omega}}
	\del_r(r^2 W_{r\phi}) - {\tau\over{\Omega}}\right) 
	- \dot{\Sigma}_W.
\end{equation}
Here $\Sigma \equiv$ surface density, $\Omega \equiv$ rotation
frequency, $\dot{\Sigma}_W \equiv$ mass lost per unit area in winds,
$W_{r\phi} \equiv \int dz\, w_{r\phi} \equiv$ height-integrated shear
stress, and $\tau \equiv$ direct torque per unit area-- possibly
supplied by a magnetohydrodynamic (MHD) wind or by the tidal field of
the secondary.  Equation (\ref{ESURF}) is in a sense fundamental; it is
derived from the angular momentum and continuity equations in the limit
that the disk is thin.

In solving equation (\ref{ESURF}) two assumptions are almost always made.
First, $\tau = \dot{\Sigma}_W = 0$.  This assumption is required by our
ignorance of the very difficult, global problem of disk winds; the
relative importance of external torques $\tau$ and internal stresses
$w_{r\phi}$ remains one of the outstanding problems of disk physics.
Second,
\begin{equation}\label{VISCDEF}
w_{r\phi} = {3\over{2}}\rho\Omega\nu; \qquad \nu = \alpha
	c_s^2/\Omega,
\end{equation}
where $c_s \equiv$ sound speed, $\nu$ is the ``anomalous viscosity''
and $\alpha$ its nondimensional counterpart.  This equation is not
fundamental; it is the simplest possible representation of the effects
of turbulence with the correct dimensional form.

So perhaps if we cannot, god-like, simulate an entire active galactic
nucleus on the computer, we can at least set the more modest goal of
using numerical experiments to understand the origin and evolution of
$w_{r\phi}$.  

The turbulent shear stress, or ``anomalous viscosity,'' has accreted an
aura of mystery over the years that is not entirely deserved.  We know
what the basic governing equations are.  They do not involve exotic
particle physics.  In many cases (ADAFs and disk coronae are
potentially important exceptions) they do not even involve plasma
kinetics.  One may simply treat the disk as a magnetized fluid.  Then
\begin{equation}
{D\bv\over{D t}} = -{\bnabla p\over{\rho}} - {\bnabla B^2\over{8\pi\rho}}
	+ {(\bB\cdot\bnabla)\bB\over{4\pi\rho}} - \bnabla \phi,
\end{equation}
where $\phi \equiv$ gravitational potential.  Any change in the angular
momentum of the fluid in the disk is due to a torque ${\bf N} = {\bf
r}\times {D\bv/Dt}$.  Torques can only be due, then, to pressure
gradients, magnetic forces, or nonaxisymmetric gravitational fields.
Radiation forces, which have been dropped here, are negligible except
in certain special circumstances.

The fluid equations are easy to write down but hard to solve.  To
make a numerical solution practical, one wants to (1) include the
minimal relevant physics, (2) take advantage of a symmetry, such as
axisymmetry, if possible, and (3) start with a simple disk model.  

The first of these considerations motivated early workers on disk
dynamics to neglect magnetic forces in the equation of motion.  Since
the discovery by Balbus \& Hawley \cite{bh91} of a linear instability
in weakly magnetized disks, it has become clear that this
approximation, while nobly motivated, misses perhaps the most important
physics (some disks, however, may be so neutral that they are decoupled
from the magnetic field \cite{g96,gm98}).

One might still hope to use axisymmetry.  Alas, early axisymmetric
models of magnetized disks showed the development of structures called
``channel solutions,'' consisting of superposed layers of material with
large radial velocities \cite{hb92}.  An analytical study
\cite{gx94} showed that these structures are unstable in three
dimensions.  The upshot is that two dimensional experiments have a
nonlinear outcome that is completely different from that of three
dimensional experiments.   While it is possible that channel solutions
may be relevant under certain special conditions, three dimensional
experiments are really required to advance our understanding of disks.

Finally, one wants to start with a simple disk model.  A natural choice
is the ``local model,'' which is a rigorous first order expansion of
the equations of motion in $\epsilon = H(r)/r \ll 1$ ($H(r) \equiv$
disk scale height) in a frame comoving with the disk.  Boundary
conditions are also needed so that the local model can be mapped to a
finite computational domain.  The ``shearing box'' boundary conditions
\cite{t81,wt88,hgb95} are well suited to this.  They are similar to
periodic boundary conditions except that they allow for the presence of
shear due to differential rotation.  They permit the study of a small
piece of the disk without reference to poorly understood inner and
outer radial boundaries, although vertical boundary conditions must
still be supplied.

All this motivates interest in numerical experiments in the shearing
box.  I will review recent work, then discuss open questions that
might be addressed with more work, and more CPU cycles, in the future.

\section*{Numerical Experiments and Results}

The 3D shearing box experiments done to date have all considered a
compressible fluid using finite difference or finite-difference-like
methods
\footnote{A incompressible study of MHD turbulence in disks using
pseudo-spectral methods is also possible and would provide an
interesting check on the numerics.
}
They have included a variety of physics: magnetic fields and pure
hydrodynamics; resistive \cite{hgb96} and ambipolar \cite{bnst95}
diffusion; other-than-Keplerian rotation curves \cite{abl96}; forced
convection from heating at the midplane \cite{sb96,c96}; and
self-gravity \cite{g98}.

Shearing box experiments may be divided into two broad classes.  The
first class are unstratified, i.e. neglect the vertical structure of
the disk.  Thus if $\phi$ is the gravitational potential, $d\phi/dz =
0$ and so in the initial laminar equilibrium $d\rho/dz = 0.$  The
second class are stratified.  They include the usual vertical tidal
potential $\phi = {1\over{2}} \Omega^2 z^2$, so an isothermal
equilibrium would have $\rho = \rho_0 \exp(-z^2/(2 H^2))$.

\subsection*{Unstratified, Magnetized Shearing Box}

Why not dispense with unstratified models and go straight to the more
realistic, stratified case?  First, the unstratified models are
numerically less demanding.  The stratified boxes include low density
regions that require small timesteps because of the Courant condition in
a magnetized fluid.  Stratified boxes also assign only a fraction of
the zones to the turbulent center of the disk; the rest are assigned to
the nearly force-free disk atmosphere.  Second, the unstratified boxes
allow the study of the nonlinear development of the Balbus-Hawley
instability independent of buoyancy effects.  A final, {\it post hoc}
justification is that stratification is not yet measured to have a
significant effect on the outcome.  This may, however, not be true in
future large, highly resolved experiments.

The model has several important dimensionless parameters.  One set of
parameters describes the shape and size of the box relative to $H =
c_s/\Omega$.  Typically $L_r \times L_\phi \times L_z = (1 \times 2\pi
\times 1) H$.  It is natural to set $L_z = (1-2)H$; this simulates the
limited vertical scale available in a disk.  The other dimensions are
limited only by CPU time.  Another set of dimensionless parameters
describes the numerical resolution.  Most experiments to date use
$2^5$ to $2^7$ zones along each axis.  A final set of parameters
describes the mean magnetic field $\<{\bf B}\>$ (the brackets denote a
spatial average), made dimensionless by comparison of the Alfv\'en speed
with $c_s$.  The boundary conditions force $\<B_z\>$ and $\<B_r\>$ to
be constant in time, while $\<B_\phi\>$ is fixed if and only if
$\<B_r\> = 0$.

Two groups have reported unstratified shearing box experiments
\cite{hgb95,mt95,hgb96}.  All show the development of MHD turbulence
initiated by the Balbus-Hawley instability; the turbulence eventually
settles down into a final state that is independent of all the initial
conditions except the mean magnetic field.

There are many interesting quantities that can be measured in the final
state.  I will focus on just one: the shear stress in the
nondimensional form
\begin{equation}
\alpha \equiv {2 w_{r\phi}\over{3\rho c_s^2}} = {2\over{3\rho
c_s^2}}\left\langle -{B_r B_\phi \over{4\pi}} + \rho v_r \delta v_\phi 
\right\rangle,
\end{equation}
where I have assumed the disk is Keplerian.  The first term is the
magnetic, or ``Maxwell,'' stress, and the second term is the fluid, or
``Reynolds'' stress.  The brackets indicate an average in $r,\phi,z$
and $t$.

The single most important result to emerge from the unstratified
shearing box experiments is that the Balbus-Hawley instability
initiates MHD turbulence that has $\alpha \gtrsim 10^{-2}$.  No other
proposed mechanism for angular momentum transport (except gravitational
instability) has produced this result in a purely local disk model.

Measured values of $\alpha$ depend on the box size, on the
mean magnetic field strength, and at worst very weakly on numerical
resolution \cite{hgb95,hgb96}.  Crudely speaking,
\begin{equation}\label{ALPHA}
\alpha \sim 0.01 + 4 {\<V_{A,z}\>\over{c_s}}
+ {1\over{4}} {\<V_{A,\phi}\>\over{c_s}},
\end{equation}
where $\<{\bf V}_A\> \equiv$ mean Alfv\'en velocity.  Thus for weak
mean fields, $\alpha \sim 10^{-2}$, while intermediate strength mean
fields raise $\alpha$ above this base level.  For mean fields that are
strong in the sense that $V_{A,i}$ is large compared to $L_i \Omega$
the field is linearly stable and equation (\ref{ALPHA}) does not apply
(see \cite{bh91,bh92} for linear stability criteria).  The flow is
laminar, and $\alpha \rightarrow 0$.  Equation (\ref{ALPHA}) should not
be taken too seriously as yet; while the sense is likely correct, the
coefficients could change substantially in future experiments.

Notice that in the absence of a mean field all the currents that
sustain the field are contained within the computational domain and so
are subject to decay.  In this sense, the zero-mean-field experiments
demonstrate the existence of a dynamo: they show that MHD turbulence in
disks, driven by the Balbus-Hawley instability, can sustain a magnetic
field in the presence of dissipation \cite{hgb96}.

Finally, some zero-mean-field experiments \cite{hgb96} included a
finite resistive diffusivity $\eta$.  These show that $\alpha$ is
sensitive to the presence of resistive diffusion.  Defining the
magnetic Reynolds number $Re_M \equiv c_s H/\eta$, turbulence decays at
$Re_M \lesssim 10^3 - 10^4$.  Dwarf novae disks in quiescence may fall
in or below this range \cite{gm98}.

\subsection*{Magnetized, Stratified Shearing Box}

Stratified shearing box experiments have been carried out by two groups
\cite{bnst95,bnst96,shgb96,abl96}.  The experiments have $L_z = $ 4-6
$\times \sqrt{2} H$, and numerical resolution similar to the
unstratified models.  Both groups report generally consistent results,
which is remarkable since very different numerical methods were used.
The stratified experiments give $\alpha \sim 10^{-2.5}$.

The stratified boxes produce $\alpha$ slightly lower than the
unstratified boxes.  Part or all of this difference may be due to the
lower effective numerical resolution of the more numerically demanding
stratified experiments.  The stratified boxes are sensitive to
resolution in that, when the resolution is increased, $\alpha$ also
increases \cite{shgb96,bnst96}.  Once higher numerical resolution
becomes practical, the stratified experiments ought to be repeated
until convergence can be demonstrated.

The measured $\alpha$ is insensitive to the vertical boundary
conditions.  Again, this may change in larger, converged experiments.
Some experiments have used vertical boundary conditions that allow the
mean field to evolve \cite{bnst95,bnst96,abl96}.  In my view this is
risky because the mean field is generated by currents ``in the
boundary'' and not within the computational volume itself; changes in
the mean field are thus generated by interaction of the fluid with the
boundary conditions.  This is, however, mainly a matter of taste.  It
is fair to say that we do not yet know what the most relevant vertical
boundary conditions are.

One of the main motivations for the stratified experiments is that they
can in principle be used to estimate the vertical run of turbulent
dissipation in disks, thereby removing a serious obstacle to predictive
models for disk spectra.  In isothermal models the run of dissipation
is not calculated directly, but two closely related quantities are:
$w_{r\phi}(z)$ and $S_z(z)$, the vertical component of the Poynting flux.
It is found that \cite{shgb96}
\begin{equation}
S_z(z = H) \simeq 0.01 \int_{0}^{H}\,dz\,{3\over{2}} \Omega \, w_{r\phi}(z).
\end{equation}
Thus only a small fraction of the power extracted by the turbulent
shear stress from the differential rotation emerges as MHD waves.
Larger values of $\alpha$ in future experiments, however, would imply
stronger fields and hence greater magnetic buoyancy.  It is also found
\cite{shgb96} that $\alpha$ is not constant with height, as is commonly
assumed.  A better fit to the data is given by $\alpha \sim
(\rho/\rho_0)^{-1/2}$, consistent with equation (\ref{ALPHA}).  Clearly
the situation with the stratified experiments is not entirely
satisfactory; it is a challenge for future experiments to make the
dissipation and vertical energy transport explicit in a converged
calculation.

\subsection*{Unmagnetized Box}

Are there any local transport processes that can compete with or
dominate MHD turbulence?  And what transport processes govern the
evolution of disks that are nearly neutral and thus poorly coupled to
the magnetic field?

Convectively driven turbulence was once thought a promising transport
mechanism \cite{c78,lp80}.  There was an early warning that something
might be amiss, however, from a quasilinear study \cite{rg92} which
showed that nonaxisymmetric convective modes produce an {\it inward}
angular momentum flux.  Subsequent shearing box experiments
\cite{sb96,c96} showed that both forced convection and overturning of
an initially unstably stratified disk led to inward angular momentum
transport ($\alpha < 0$) in the fully nonlinear regime.  This
counterintuitive result is a nice illustration of the value of
numerical experiments.

There is still one regime in which convectively driven angular momentum
transport could play a role.  That is in geometrically thick flows with
unstable radial stratification-- for example ADAFs (see \cite{nmq97}
for a review).  A shearing box experiment with forced radial convection
might be revealing, but the shearing box is not a good model for these
flows.  Global models are really required.

Shearing box experiments have also permitted the direct evaluation of
another once-promising mechanism for initiating turbulence in disks:
nonlinear hydrodynamic instability \cite{bhs96}.  In these experiments a
purely hydrodynamical Keplerian shear flow is violently perturbed.  It
is found that the flow returns to a laminar state, independent of the
amplitude of the initial perturbation.

Because the numerical experiments are run at a Reynolds number that is
low in comparison to that in astrophysical disks, nonlinear
hydrodynamic instability cannot be rigorously ruled out.  But analytic
arguments \cite{bhs96} (see also Balbus, this volume) and the absence
of nonlinear instability in laboratory analogs relevant to Keplerian
shear flow (see \cite{bh98} and references therein) make it likely, in
my view, that Keplerian disks are nonlinearly hydrodynamically stable.

Finally, self-gravity may be noticeable or even dominant in disks
around young stars and in active galactic nuclei.  The shearing box has
been used to study gravitational instability in cooling disks
\cite{g98}.  Cooling is essential since in its absence the disk simply
heats up until it is stable.  The outcome is a fluctuating state with
Toomre's stability parameter $\<Q\> \sim 1$ and significant outward
transport of angular momentum via gravitational and Reynolds stresses.
These experiments also show that self-gravity produces truly local
transport in a sense to be made clear below.

\section*{Open Questions}

I will conclude with a short list of open questions that are
astronomically relevant and can be answered in the near future with
numerical experiments in the shearing box.

1.  {\it Is $\alpha$ determined locally?}  A different way of phrasing
this question is, does $\alpha$ converge as the planar box size $L_r$
and $L_\phi$ are increased?  Experiments to date find turbulence with
most of the energy in structures with scale comparable to the box
size.  Thus the outcome is limited by the box size.  

If $\alpha$ is determined locally one expects that the autocorrelation
function of fluid variables to decay rapidly on scales $\gtrsim H$.
Equivalently, fluid variable power spectra should turn over and decline
at small $k_r$ and $k_\phi$.  Because the natural scale for the
turnover is $H$, it should be seen in boxes only slightly larger than
current experiments.  The detection of the turnover would be a
significant milestone for disk physics in that it would show that
$\alpha$ is locally determined and so validate the use of the local
model.

The alternative is that the fluid autocorrelation functions decline
only slowly or not at all with distance in the plane of the disk.  Then
the shear stress at any point in the disk can be influenced by
conditions at points many scale heights away.  This is not
inconceivable.  The $\alpha \sim (H/r)^n$ prescription common in
studies of dwarf novae, for example, requires this sort of nonlocality
because local turbulence must somehow ``know'' about the large scale
structure of the disk and in particular the local radius $r$.  If
$\alpha$ is nonlocal, however, we must abandon the shearing box and
move to global models.  In what follows I assume that $\alpha$ is local
and thus that the shearing box is relevant.

2. {\it How does $\alpha$ depend on resistive diffusion $\eta$, viscosity
$\nu$, and their ratio, the magnetic Prandtl number $Pr_M \equiv
\nu/\eta$?} In most circumstances both resistivity and viscosity are
negligible, but even then, it has been argued, $Pr_M$ may govern the
character of MHD turbulence (see \cite{bh98} and references therein).
This hypothesis can be tested with sufficiently high resolution
experiments that allow a reasonable separation of the resistive and
viscous lengthscales.  Resistivity is not always negligible, however.
Recent work \cite{gm98} shows that the standard disk instability model
for dwarf novae implies a magnetic Reynolds number of order $10^3$ for
dwarf nova disks in quiescence.  Such low $Re_M$ may have a direct
impact on the development of MHD turbulence (cf. \cite{hgb96}).
Protostellar disks also suffer from high resistivity; some parts are
likely to be completely decoupled from the magnetic field \cite{g96}.
This issue has direct astronomical relevance and can be studied with 
codes and computers that are now available.

3. {\it How does $\alpha$ vary in time?}  Local model MHD experiments
have so far only sought to measure $\alpha$ when the disk scale height
is steady in time.  Some of the most potentially revealing phenomena in
disk systems, however, involve rapid changes in disk temperature.
Under these circumstances, how long does it take $\alpha$ to readjust
to its steady state value?  There is already preliminary evidence that
this relaxation time is long in that the time required for the
zero-mean-field unstratified experiments to reach a steady state is
many orbital periods \cite{hgb96}.  But a more direct determination can
be made in stratified, isothermal shearing box experiments in which the
temperature is forced to change suddenly.  If the relaxation time
is long compared to the thermal timescale, then the structure of
cooling and heating fronts in dwarf nova disks may be quite different
from what is now imagined.

4. {\it How does $\alpha$ vary near disk edges?}  Gaps may be opened in
disks around young stellar objects by stellar or planetary companions.
The size of the gap and accompanying torque on the companion are
sensitive to the surface density profile close to the disk edge, which
is determined by a balance between tidal torques on the disk and all
other torques.  One way of studying disk edges would be to allow a
point mass to orbit in the shearing box and clear a gap in the disk.
Such an experiment could test whether $\alpha$ varies near the edge of
the disk in a way that is consistent with the usual treatment of
$w_{r\phi}$ as a viscous stress.  Perhaps it does not; perhaps $\alpha$
increases rapidly within a few $\times 2\pi H$ of the edge.  Such
experiments could also test whether magnetic fields drive cross-gap
accretion.  Numerical work to date has focused on the nonmagnetic
problem (e.g. \cite{al96}), but this approximation may be missing an
important piece of the puzzle.

\section*{Acknowledgements}

I am grateful to Steve Balbus, John Hawley, Kristen Menou, Eve
Ostriker, Eliot Quataert, and Jim Stone for their comments.  This work
was supported by NASA grant NAG 52837.


\begin{references}

\bibitem{o85}
Ostriker, J. P. 1985, in The Milky Way Galaxy, eds. H. Van Woerden, R.
J.  Allen, \& W. B. Burton (Dordrecht: Reidel) 635

\bibitem{c93}
Cannizzo, J. 1993, in Accretion Disks in Compact Stellar Systems
ed. J. Wheeler (Singapore: World Scientific), 6

\bibitem{hmdlh98}
Hameury, J.-M. et al. 1998, in preparation

\bibitem{bh91}
Balbus, S.A., \& Hawley, J.F. 1991, ApJ, 376, 214

\bibitem{g96}
Gammie, C.F. 1996, ApJ, 457, 355

\bibitem{gm98}
Gammie, C.F., \& Menou, K. 1998, ApJ, 492, L75

\bibitem{hb92}
Hawley, J.F., \& Balbus, S.A. 1992, ApJ, 400, 595

\bibitem{gx94}
Goodman, J., \& Xu, G. 1994, ApJ, 432, 213

\bibitem{t81}
Toomre, A. 1981 in The Structure and Evolution of Normal Galaxies,
eds. S.M. Fall \& D. Lynden-Bell (Cambridge: Cambridge University
Press), 111

\bibitem{wt88}
Wisdom, J., \& Tremaine, S. 1988, AJ, 95, 925

\bibitem{hgb95}
Hawley, J.F., Gammie, C.F., \& Balbus, S.A. 1995, ApJ, 440, 742

\bibitem{hgb96}
Hawley, J.F., Gammie, C.F., \& Balbus, S.A. 1996, ApJ, 464, 690

\bibitem{bh92}
Balbus, S.A., \& Hawley, J.F. 1992, ApJ, 400, 610

\bibitem{bnst95}
Brandenburg, A., Nordlund, A., Stein, R.F., \& Torkelsson, U. 1995,
ApJ, 446, 741

\bibitem{abl96}
Abramowicz, M., Brandenburg, A., \& Lasota, J.-P. 1996, MNRAS, 281, L21

\bibitem{sb96}
Stone, J.M., \& Balbus, S.A. 1996, ApJ, 464, 364

\bibitem{c96}
Cabot, W. 1996, ApJ, 465, 874

\bibitem{g98}
Gammie, C.F. 1998, in preparation

\bibitem{mt95}
Matsumoto, R., \& Tajima, T. 1995, ApJ, 445, 767

\bibitem{bnst96}
Brandenburg, A., Nordlund, A., Stein, R.F., \& Torkelsson, U. 1996,
ApJ, 458, L45

\bibitem{shgb96}
Stone, J.M., Hawley, J.F., Gammie, C.F., \& Balbus, S.A. 1996, ApJ, 463, 656

\bibitem{c78}
Cameron, A.G.W. 1978, Moon Planets 18, 5

\bibitem{lp80}
Lin, D.N.C., \& Papaloizou, J.C.B. 1980, MNRAS, 191, 37

\bibitem{rg92}
Ryu, D. \& Goodman, J. 1992, ApJ, 388, 438

\bibitem{nmq97}
Narayan, R., Mahadevan, R., \& Quataert, E. 1997, in
Proc. Reykjavik Symp. on Non-linear Phenomena in Accretion Disks Around
Black Holes, eds. M.A. Abramowicz, G. Bjornsson, \& J.E. Pringle,
in press

\bibitem{bhs96}
Balbus, S.A., Hawley, J.F., \& Stone, J.M. 1996, ApJ, 467, 76

\bibitem{bh98}
Balbus, S.A., \& Hawley, J.F. 1998, Rev. Mod. Phys., in press

\bibitem{al96}
Artymowicz, P., \& Lubow, S.H. 1996, ApJ, 467, L77

\end{references}
\end{document}